\documentclass[aps,twocolumn,groupedaddress,showpacs,floatfix]{revtex4-stper}
\usepackage{dcolumn}
\usepackage{graphicx}
\usepackage{natbib}
\usepackage{epstopdf}
\bibpunct{[}{]}{,}{n}{}{}

\begin{document}
\title{Applying a resources framework to analysis of the Force and Motion Conceptual Evaluation}
\author{Trevor I. Smith and Michael C. Wittmann}
\affiliation{Department of Physics and Astronomy, University of Maine, Orono, ME 04469}

\begin{abstract}
We suggest one redefinition of common clusters of questions used to analyze student responses on the Force and Motion Conceptual Evaluation (FMCE). Our goal is to move beyond the expert/novice analysis of student learning based on pre-/post-testing and the correctness of responses (either on the overall test or on clusters of questions defined solely by content). We use a resources framework, taking special note of the contextual and representational dependence of questions with seemingly similar physics content. We analyze clusters in ways that allow the most common incorrect answers to give as much, or more, information as the correctness of responses in that cluster. Furthermore, we show that false positives can be found, especially on questions dealing with Newton's Third Law.
\end{abstract}
\maketitle

\section{Introduction}
Students are not yet physicists. They have not had the extensive training that we, as physicists, rely on. As a result, it is often inappropriate to categorize and group student responses to physics questions solely on the basis of agreeing with correct Newtonian principles, as is commonly done with standardized tests such as the Force Concept Inventory (FCI) and the Force and Motion Conceptual Evaluation (FMCE).\cite{fci,hake,thornsok} Unfortunately student assessment using the FMCE, including work previously done by one author (M.C.W.), regularly does just that.\cite{wittmannfmce} The test questions are grouped into clusters according to a physicist's view of equivalent content areas, and students' responses are evaluated based on their agreement with a physicist's viewpoint without regard for why students might choose incorrect answers. This may be a valid form of assessment for determining how well students think like physicists, but it is often an insufficient method for determining how students reason about scenarios in a physics context. In order to effect greater conceptual development in our students, we must understand not only where we wish them to end up, but also where they are beginning in terms of their understanding of the world around them. Only by having this entire picture may we devise a manner by which to help our students truly gain a physicist's understanding of their surroundings.

This paper describes ways in which the FMCE can be organized and used to get more detailed information about students. Presently, many researchers and educators using the FMCE follow a particular procedure that includes three steps: 1) administering the FMCE pre- and post-instruction, 2) dividing the questions into content-based clusters, and 3) evaluating the correctness of each student's responses within each cluster as well as over the entire test. Several years ago a template was developed by one author (M.C.W.) to analyze students' responses to the FMCE within five clusters (Velocity, Acceleration, Force (Newton's $1^{st}$ and $2^{nd}$ laws), Newton's $3^{rd}$ law\cite{MBL1}, and Energy).\cite{wittmannfmce} The template automatically scores each response as correct or incorrect, groups questions into the aforementioned clusters, and calculates a class's normalized gain for each cluster as well as over the entire test. This template has become widely used due to its availability and its succinct analysis of students' responses. Recent research using the FMCE and modeling using a resources framework\cite{Hammer2000, minstrell1992, wittmann2006} has convinced us that analysis based on the use of this template lacks the depth and rigor we have come to expect from studies on students' understanding of Newtonian mechanics. We propose several modifications to the described analysis including a redefinition of clusters and a deeper analysis of students' incorrect responses.

The clusters mentioned previously divide the FMCE very nicely into groups of questions that each examine a different content aspect of physics.  But many studies, including those by Beichner\cite{tugk} and Kohl and Finkelstein\cite{kohl,kohl2}, have shown that the manner in which material is presented may significantly affect students' abilities to demonstrate their understanding. For example, Beichner has shown that students' understanding (or lack of understanding) of graphs can have a profound impact on their responses to physics questions involving graphs.\cite{tugk} Furthermore, results by Dykstra and others show that elements of the physical situation greatly affect reasoning. Dykstra reports on students' troubles reasoning about motion in which an object has a turning point; that is, when an object under the influence of a constant force moves in a particular direction while slowing down and then reverses direction and speeds up.\cite{dykstra}

Our goals in this paper are not new. Several researchers (notably Thornton\cite{thorntonblue,thorntonwhite} and Dykstra\cite{dykstra}) have used results from the FMCE to give fine-grained analyses of students' responses to the FMCE.  Unfortunately, their methods are not widely used among physics educators and education researchers. Also, we wish to anchor our analysis in a resources framework, making explicit connections to the representational and contextual dependence of responses. In addition, Bao has proposed new clusters of questions to investigate reasoning. We respond to Bao's work in more detail in section \ref{sec:newclusters}. 

In section \ref{sec:newclusters} we both respond to existing clustering methods (including our previous one) and propose new clusters. In section \ref{sec:resources} we discuss various incorrect mental models that correspond with the content areas described by each of our clusters.  In section \ref{sec:models} we examine how these mental models are aligned with particular responses to questions in the FMCE. Using a definition of clusters consistent with a resources framework allows us to go into greater detail about students' incorrect mental models and identify responses that correspond to these models. In this section, we include a discussion of false positives, as measured by looking at responses on several questions within a representationally and contextually consistent set of questions.

\section{Revised Question Clusters for the FMCE}
\label{sec:newclusters}
The five analysis clusters, shown in table \ref{clusters_old}, were chosen by Wittmann as a quick and dirty analysis of classroom performance on the FMCE.  These clusters were defined based on the content of each question on the FMCE---the Velocity cluster asks students about the velocity of an object undergoing a series of described motions, the Force (Newton I \& II) cluster asks students about the force (s)  exerted on an object during a described motion, and so on.  Several questions are not included in any of the clusters; Thornton and Sokoloff omit these from regular analysis of the FMCE because they are intended for diagnostic purposes (such as reading ability) or  do not give a definitive indication as to whether or not a student is properly using Newtonian reasoning.\cite{thornsok}  More details of these omissions can be found in refs. \cite{thornsok2, thorntonwhite}. 

\begin{table}[ht]
\caption[FMCE: Old Question Clusters]{Clusters of questions on the FMCE as previously defined by Wittmann.}
\begin{center}
\begin{tabular}{cc}
\hline\hline
Cluster&Questions\\
\hline
Velocity&40--43\\
Acceleration&22--29\\
Force (Newton I \& II)&1--4, 7--14, 16--21\\
Newton III&30--32, 34, 36, 38\\
Energy&44--47\\
\hline\hline
\end{tabular}
\end{center}
\label{clusters_old}
\end{table}%

There are obvious flaws to the old clustering of questions, the largest being that different representations and contexts are asked about in many clusters. If students are inconsistent in their thinking about the physics (as assumed in a resources framework), then results in each cluster should be noisy and inconsistent, as well. 

\subsection{Defining new clusters}
\label{subsec:clusterdefine}
\begin{figure}[ht]
\centering
\includegraphics[totalheight=3.5in]{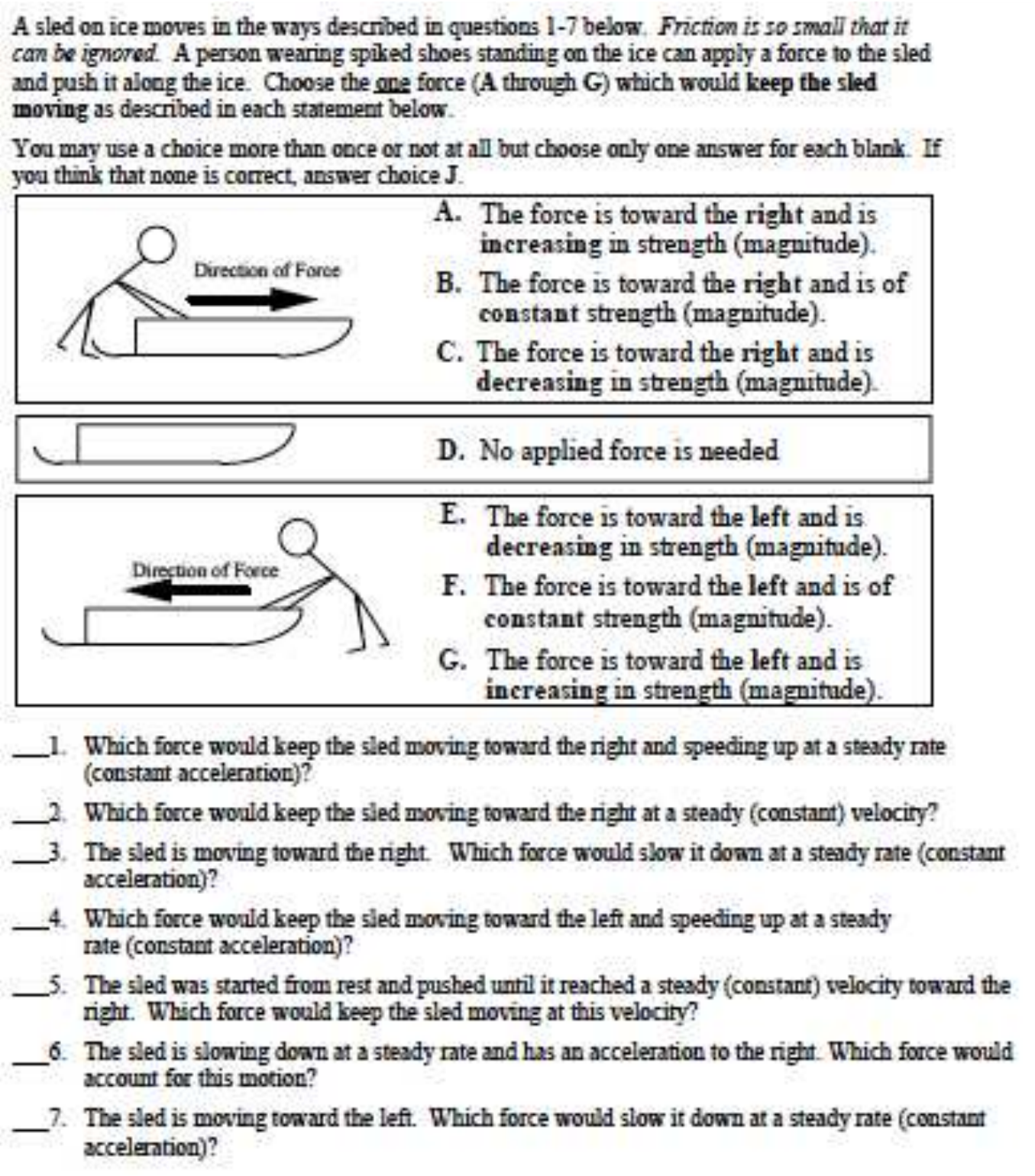}
\caption{Questions 1--7 of the FMCE\cite{thornsok}:  contained within original Force (Newton I \& II) cluster and revised Force Sled cluster.}
\label{force_sled_ex}
\end{figure}

\begin{figure}[ht]
\centering
\includegraphics[totalheight=4in]{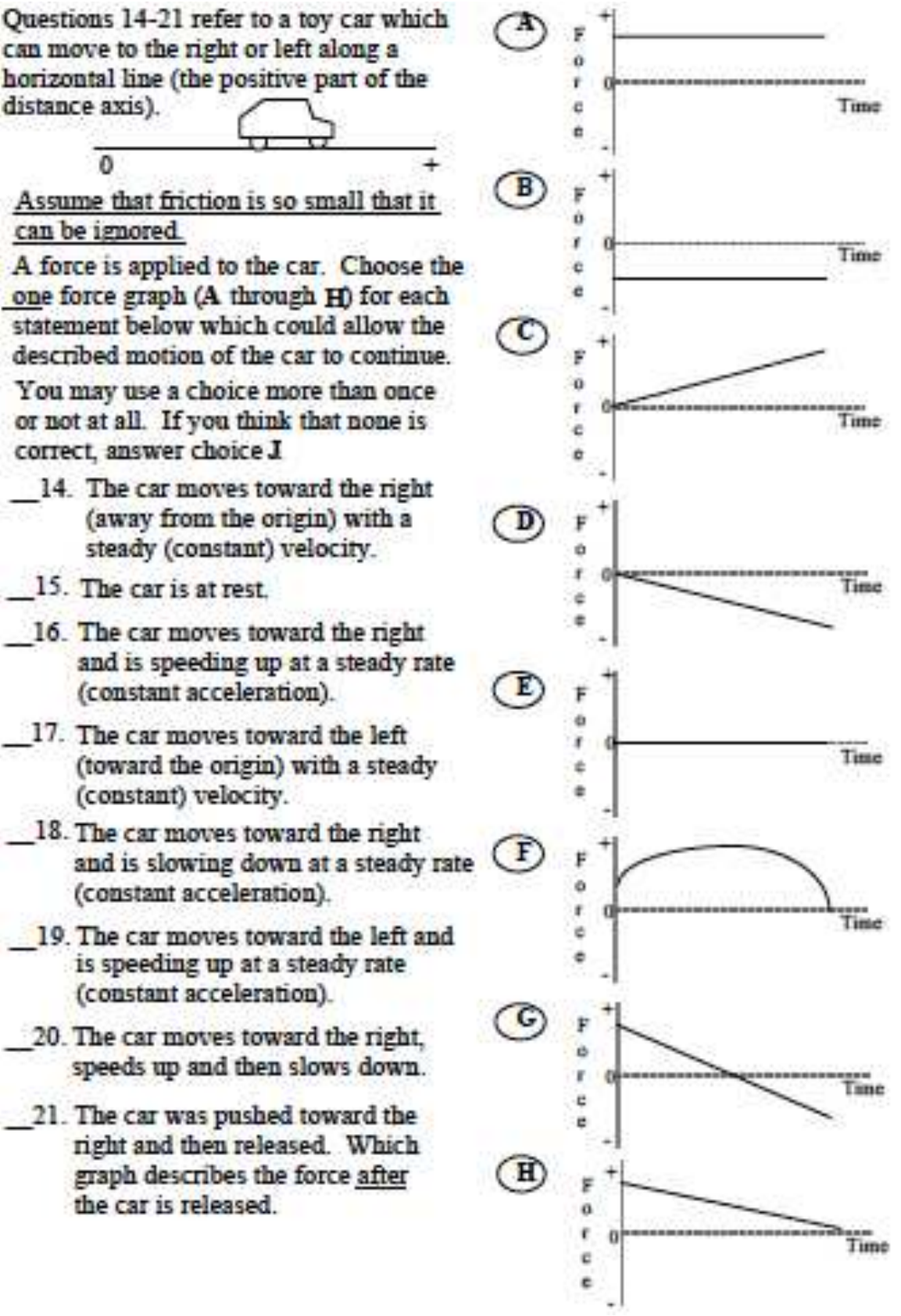}
\caption{Questions 14--21 of the FMCE\cite{thornsok}:  contained within original Force (Newton I \& II) cluster and revised Force Graphs cluster.}
\label{force_graphs_ex}
\end{figure}

\begin{figure}[ht]
\centering
\includegraphics[totalheight=3.5in]{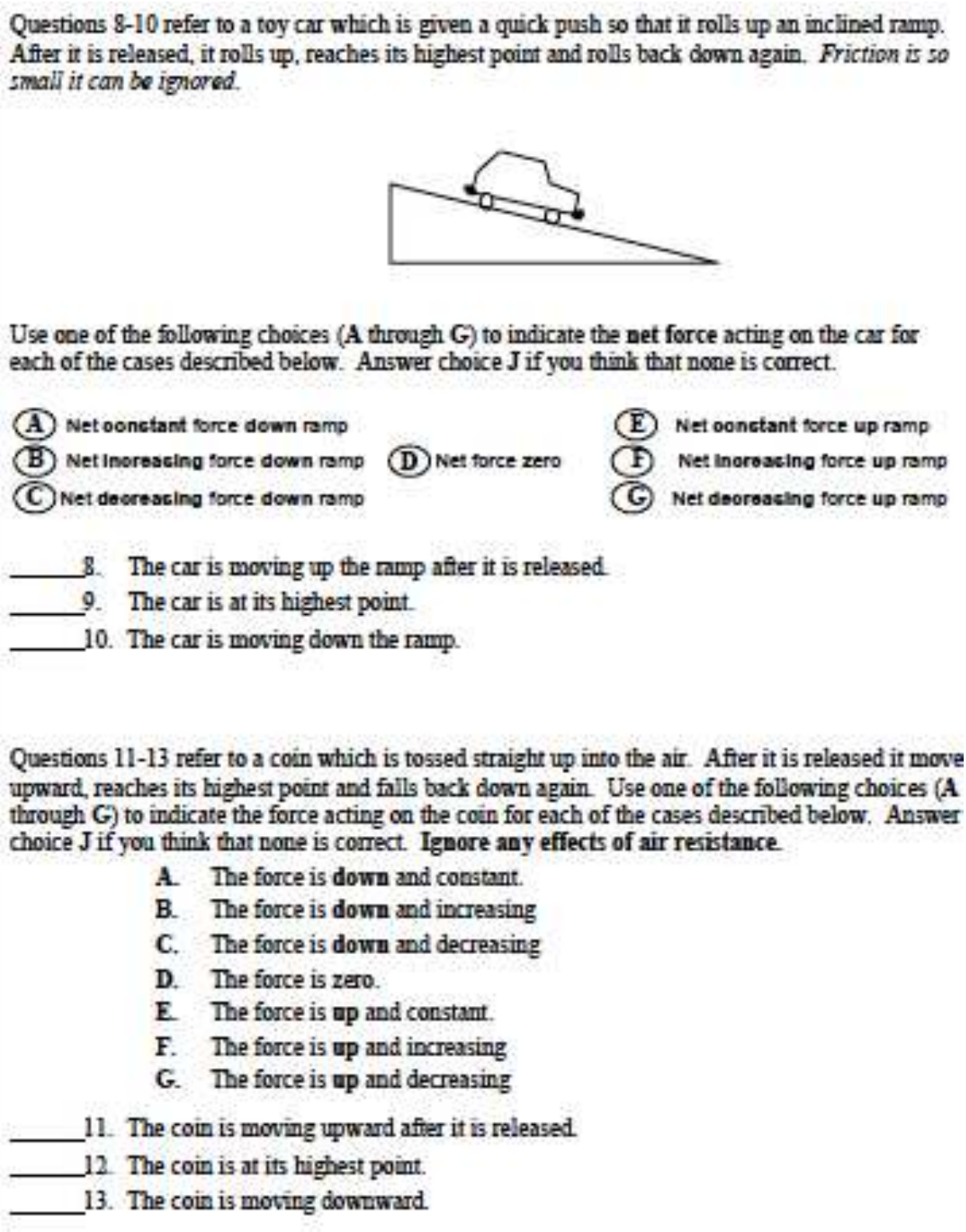}
\caption{Questions 8--13 of the FMCE\cite{thornsok}:  contained within original Force (Newton I \& II) cluster and revised Reversing Direction cluster.}
\label{vertical_ex}
\end{figure}

To understand student reasoning better, we should use a finer-grain resolution in the questions that we analyze and group. We use the questions of the Force (Newton I \& II) cluster, shown in figures  \ref{force_sled_ex}--\ref{vertical_ex},  to illustrate.

In answering the questions in figures \ref{force_sled_ex}--\ref{vertical_ex}, students must determine the force on an object (sled, toy car, coin) undergoing a described motion.  In terms of representations, figures \ref{force_sled_ex} and \ref{vertical_ex} use pictorial representations, while figure \ref{force_graphs_ex} uses graphical representations. In terms of context,  the questions in figure \ref{force_sled_ex} and \ref{force_graphs_ex} are about motion in a single direction, while the questions in figure \ref{vertical_ex} involve an object that reverses the direction. 

To measure dependence of student reasoning on the representational and contextual cues  in figures \ref{force_sled_ex}--\ref{vertical_ex}, we have created new clusters which replace the old Force (Newton I \& II) cluster:  the Force Sled cluster (containing the questions in figure \ref{force_sled_ex}), the Force Graphs cluster (containing the questions in figure \ref{force_graphs_ex}), and the Reversing Direction cluster (containing the questions in figure \ref{vertical_ex} as well as others).  The Reversing Directions cluster has been expanded to include questions about acceleration as well as force; questions 27--29 on the FMCE inquire about the acceleration of a coin tossed in the air as it moves up and back down, isomorphic to questions 11--13 but in the context of acceleration.

Table \ref{clusters_new} shows the full definitions of the revised clusters, ordered by question number on the test rather than difficulty of the physics material.  These clusters are consistent with those used by Thornton.\cite{thorntonblue} Note that the definitions of the original Newton III, Velocity, and Energy clusters have been directly transferred to the revised clusters.  To conform with the specificity of the Force Graphs and Acceleration Graphs clusters, the Velocity cluster has been renamed the Velocity Graphs cluster.  Most of the original Acceleration cluster has been transferred to the Acceleration Graphs cluster (the other acceleration questions are in the Reversed Direction cluster). These revised clusters increase the information that can be taken from an analysis of the FMCE by highlighting and isolating content areas, types of representations, and specific situations with which students may struggle. We give examples in section \ref{sec:models}.

\begin{table}[ht]
\caption[FMCE: Question Clusters]{Revised clusters on the FMCE.}
\begin{center}
\begin{tabular}{ccc}
\hline\hline
Cluster&&Questions\\
\hline
Force Sled&&1--4, 7\\
Reversing Direction&&8--13, 27--29\\
Force Graphs&&14, 16--21\\
Acceleration Graphs&&22--26\\
Newton III&&30--32, 34, 36, 38\\
Velocity Graphs&&40--43\\
Energy&&44--47\\
\hline\hline
\end{tabular}
\end{center}
\label{clusters_new}
\end{table}%

\subsection{Comparing to other clusters}
\label{subsec:clusterresponse}

Other ways of clustering questions on the FMCE exist.  In his Ph.D. dissertation, Bao claims that mixed model states are more easily detected using samples of questions that span several physical contexts.\cite{baodiss}  As such he has defined the cluster of questions in table \ref{baocluster} to compare students' use of two particular models (force proportional to acceleration, and force proportional to velocity).\footnote{Table \ref{baocluster} also shows the response to each question that corresponds with each of these models.}  His cluster contains questions from both the Force Sled (question 2) and Reversing Direction (questions 11 \& 12) clusters as well as question 5, which Thornton and Sokoloff suggest should only be used as a measure of reading ability rather than Newtonian thinking.\cite{thornsok2}  

Using a resources framework, we can explain the presence of mixed model states as being the result of contextual cues or representational cues. Thus, by creating a cluster that mixes cues, he has primed his data to show mixed model states while not being able to explain the source of model mixing, be it contextual, representational, or due to deeper issues with the physics. Of course we expect our students to understand concepts in all contexts, but mixing cues makes our analysis of student reasoning much more difficult and fails to give the resolution that we, as researchers and instructors, desire. 

A further weakness of the mixed context cluster is the use of question 5, which describes two kinds of motion (accelerating followed by constant velocity), and should be checked against question 2 on the FMCE for consistency. Typically, students score very well on this question (unless they have reading problems in understanding the question). Thus, adding this question necessarily skews the data toward correct model use.

We believe that it is far more beneficial for instructors to first group questions that deal with a single physical context (e.g.~constant force applied to move an object horizontally) before combining questions across diverse contexts.  Our approach is more consistent with the assumptions of a resources framework, and gives new insight (as described below in the section on false positives).  We do still observe students' use of mixed models, but within a single physical context. Such a mixed model indicates a kind of inconsistency in thinking about the physics that a mixed-context cluster cannot.

\begin{table}[htdp]
\caption{Question cluster defined by Bao to use Model Analysis with the FMCE.\cite{baodiss}}
\begin{center}
\begin{tabular}{cccc}
\hline\hline
\multicolumn{1}{p{1.5cm}}{\centering{Question Number}}&\raisebox{-1.25ex}[0pt]{~~$F\propto \frac{\Delta v}{\Delta t}$~~}&\raisebox{-1.25ex}[0pt]{~~$F\propto v$~~}&\raisebox{-1.25ex}[0pt]{Other}\\
\hline
2&D&B&others\\
5&D&B&others\\
11&A&G&others\\
12&A&D&others\\
\hline\hline
\end{tabular}
\end{center}
\label{baocluster}
\end{table}

\section{Facets, Resources and Mental Models of student reasoning on the FMCE}
\label{sec:resources}
As we have stated from the outset, we use a resources framework to cluster responses on the FMCE. We are not using a conceptions approach\cite{wandersee1994,vosniadou,thorntonwhite} because we are seeking a higher resolution to understand what students have mastered in their physics learning (and how best to help those who have not yet mastered the material). The resources framework can be thought of as a schema theory that emphasizes knowledge-in-pieces, such as phenomenological primitives (p-prims)\cite{disessa1993}, facets\cite{minstrell1992}, and resources\cite{hammertransfer,wittmann2006}.  The differences and connections between a concept-based and resource-based analysis of student reasoning and teaching are discussed in more detail by Scherr.\cite{scherr2007}   

Student thinking is rarely described in terms of an individual, small idea such as ``closer means stronger", for which appropriate applications are sitting by a fire to be warmer or moving from speakers at a concert to save one's hearing, and false applications include attributing the change in seasons to a difference in the distance from the earth to the sun. Instead, succinct descriptions of student thinking often require us to recognize the use of multiple resources in connection. We represent this as a type of nodal mental space network, a resource graph.\cite{wittmann2006} The connections, or links, between these resources vary greatly in both strength and duration. Assuming that students are using a set of resources related to mechanics and kinematics, not all activated in every question, we may examine how various resources could combine to create robust (and often incorrect) mental models (or concepts) that students use when reasoning about physics. We give several examples below.

In this section, we describe the resources that students often use when answering questions on the FMCE. In some cases, we draw resource graphs. In section \ref {sec:models}, we connect these descriptions of resources to the questions on the FMCE.

\subsection{Dynamics}

\subsubsection{Newton's 1st and 2nd laws}
The notion that the force exerted on an object is proportional to its velocity has been reported by many researchers and is very prevalent among physics students.\cite{thorntonwhite,brownclement,hallhes2} The $F \propto v$ model has been described as being similar to the Impetus view of physics\cite{hallhes2} but can be described in more detail by several of Hammer's resources including ``activating agency" and (in particular) a ``maintaining agency"\cite{Hammer2000} that is ``dying away."\footnote{These and many other resources are lightly derived from diSessa's p-prims.\cite{disessa1993}} The ``activating agency" resource is the notion that every event must have a cause, i.e.~every object that is in motion must have had something to get it started. The ``maintaining agency" resource embodies the idea that objects in motion must have something (some ``agent") to keep them in motion. While neither of these resources is incorrect in and of itself, the $F \propto v$ model is evident when the ``agent" required for each of them is seen as the force applied to an object.\footnote{One could argue that this is only a problem if the students use the \textit{net} force for each of these resources; however, the fact that the FMCE presents questions that involve a single applied force on a frictionless surface makes this a moot point.} The ``maintaining agency" when used in this context contradicts Newton's first law, but students' intuitive ideas are not unreasonable. They are, instead, consistent with years of experience in our friction-filled world where a continuous application of force is almost always needed to maintain an object's motion.

\subsubsection{Newton's 3rd law}
Studies have shown that students use a variety of strategies when reasoning about the forces exerted between two interacting bodies.\cite{baohoggzoll,smith,maloney} Bao, Hogg, and Zollman identified four ``contextual features" that students use when responding to questions regarding Newton III: velocity, mass, pushing and acceleration.\cite{baohoggzoll}  For example, an object with a larger initial speed will exert more force than an object with a smaller initial speed (during a collision), and more massive objects exert more force than less massive ones.  The velocity and mass features work well together to illuminate students' implicit confusion between momentum and velocity.  Based on their ideas about kinematics, students often have a desire to represent force as $F=mv$;\cite{hallhes2} furthermore, students may use the terms momentum and force interchangeably.\cite{baohoggzoll}  The pushing feature is contained in the notion that, when one object pushes another, the object that is pushing must exert more force than the object that is being pushed.  This idea is typically accompanied by the reasoning that if both objects exerted the same force on each other, neither would move.

We have previously reported on students' use of three facets when considering Newton's third law: the mass dependence facet, the action dependence facet, and the velocity dependence facet.\cite{smith}  These mental models correspond with Minstrell's facets of knowledge,\cite{minstrell1992} in particular facets 62 (The moving object or a faster moving object exerts a greater force), 63 (The more active or energetic object exerts more force), and 64 (The bigger or heavier object exerts more force).\footnote{Definitions of these facets can be found in refs.\cite{smith,minstrell1992}.}  The mass dependence facet has a direct correlation with the mass contextual feature described by Bao, et al.\cite{baohoggzoll} The action dependence facet combines the velocity and pushing contextual features described above to create a mental model that is applicable to both pushing and collision situations.  The velocity dependence facet describes students' use of force as an intrinsic property of an object, similar to momentum, agreeing with the velocity contextual feature described by Bao, et al.\cite{smith,baohoggzoll} Maloney uses many similar ideas to describe students' use of a ``dominance principle" to reason about the interaction between two bodies.\cite{maloney}  Maloney's ``dominance principle" is also very closely related to diSessa's ``Ohm's p-prim" as well as Hammer's ``more is more" resource.\cite{disessa1993,Hammer2000}

The ``more is more" resource might manifest itself within students' thinking of Newton's third laws as a series of connections: the more mass or speed an object has,\footnote{Possibly tacit indicators of momentum or kinetic energy.} the more damage it can do; the more damage an object can do, the more force it must exert on any other object.  Another connection that can be made along these lines is the idea that the more an object reacts after a collision, the more force must have been applied to it.\cite{Elby1}  In this case, the ``more is more" resource is used indirectly to describe the object that is exerting the force on the object in question.  

\begin{figure}[ht]
\centering
\includegraphics[totalheight=4.25cm]{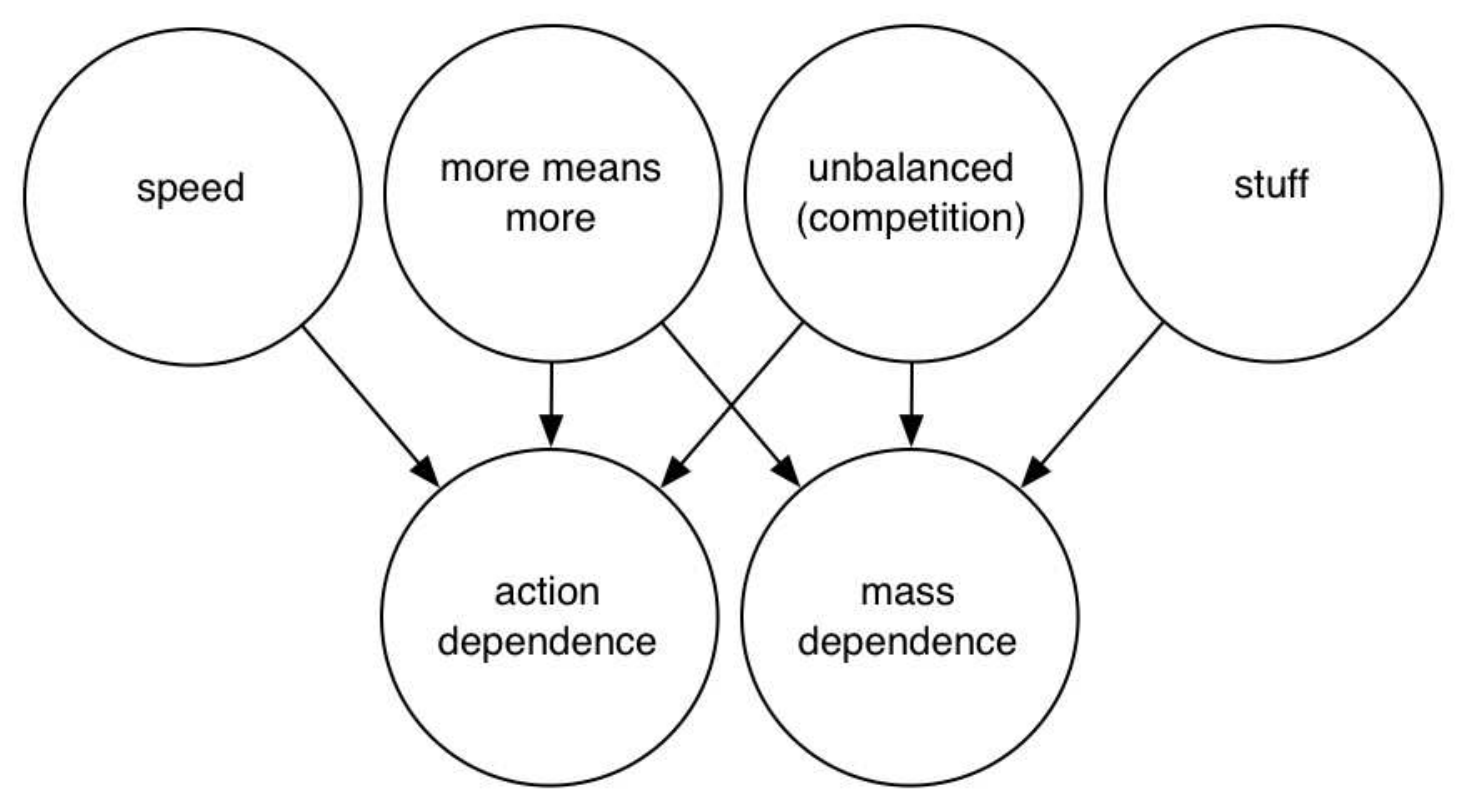}
\caption{The mass dependence and action dependence resources may be derived from universal primitives and observations of a scenario.}
\label{n3_resources}
\end{figure}

We draw a resource graph of student reasoning about Newton's third law in order to summarize these comments. In figure \ref{n3_resources} we show how four resources can combine in groups of three to create the two observed mass and action dependencies. Note that the combination of ``More means more" and ``Unbalanced (competition)" can be interpreted as diSessa's Ohm's p-prim\cite{disessa1993}, in which more resistance (say, a mass in the way) requires more force for equal effect. 

\subsection{Energy}
The ``more is more" resource discussed above may also be applied when discussing students' views of the transfer of energy.  For this particular discussion we will use the scenario of a sled starting from rest at the top of an icy hill and sliding all the way down (as is seen in the FMCE).  In this case, the ``more is more" resource may be used quite nicely (and correctly) in stating that the more height a hill has, the more kinetic energy (and thereby speed) the sled will have once it has reached the bottom.  Students may connect ``more is more" to other elements of the problem, instead, including more steepness or more length. Students might, for example, take the approach that the steeper a hill is, the faster (or more energetic) the sled will be when it gets to the bottom.  Students most likely take this idea from their own experiences sliding down hills; the steeper hills are always more fun and get them to high speeds sooner.  They attach the ``more is more" resource to the acceleration of the sled, i.e.~the rate of change of the velocity varies with the slope of the hill but not the total change in speed.  

\section{Mental Models Evident in Each Cluster}
\label{sec:models}
We return to the revised clusters defined in section \ref{sec:newclusters}, applying the resources presented in section \ref{sec:resources}.  For each cluster of questions, we examine the possible responses to individual questions and determine which correspond with the use of correct Newtonian reasoning and which indicate the use of one of the mental models discussed in section \ref{sec:resources}.  We will also compare these responses with the most common student responses reported by Thornton,\cite{thorntonblue} showing that the FMCE can be interpreted in ways consistent with a resources framework. Furthermore, we will show that the use of a resources framework lets us conclude that some student responses are actually false positives (i.e.~correct responses given for incorrect reasons).

\subsection{Force Sled Cluster}
The Force Sled cluster (see figure \ref{force_sled_ex}) asks questions in plain language (i.e.~not graphically), has no reversing direction questions, and deals with a single applied force that is therefore also the net force on the sled. Offered responses on this cluster include the correct idea that the net (and applied) force is proportional to its acceleration (or rate of change of velocity), as well as the notion that the net force on the sled is proportional to its velocity.  Table \ref{fmce_sled} shows the responses that correspond with each of these models as well as the most common student responses. The most common student responses found by Thornton\cite{thorntonblue} are the same as those indicating a student's use of the $F \propto v$ model.  This similarity provides strong evidence that many students believe that the net force on an object is proportional to its velocity, rather than its acceleration. We interpret these results in terms of resource activation, though this interpretation is, in this case, not necessary.

\begin{table}[ht]
\caption[FMCE: Force Sled cluster]{The ``Force Sled" cluster on the FMCE.}
\begin{center}
\begin{tabular}{cccc}
\hline\hline
Question&&&Most Common\\
Number&\raisebox{1.5ex}[0pt]{$F\propto\frac{\Delta v}{\Delta t}$}&\raisebox{1.5ex}[0pt]{$F\propto v$}&Student Response\cite{thorntonblue}\\
\hline
1&b&a&a\\
2&d&b&b\\
3&f&c, g&c\\
4&f&g&g\\
7&b&e&e\\
\hline\hline
\end{tabular}
\end{center}
\label{fmce_sled}
\end{table}

\subsection{Reversing Direction Cluster}
The Reversing Direction cluster (see figure \ref{vertical_ex}) asks questions in which an object has been tossed in the air (or rolled up a hill). Students must think about the net force and the acceleration throughout its up-and-down, free-fall motion.\footnote{Note that, while the ramp on which the toy car rolls in questions 8--10 prevents it from truly being in free-fall, the up and down motion of the car travelling under its own volition is analogous to the coin toss in questions 11--13.\cite{thorntonblue}}  Within the Reversing Direction cluster, the questions are broken into three sub-clusters (8--10, 11--13, and 27--29), each of which involves a single object undergoing an up-and-down motion.  In each of these sub-clusters the students are asked about the force exerted on the object (questions 8--13) or the acceleration of the object (questions 27--29) as it goes up (questions 8, 11, and 27), at the highest point in its journey (questions 9, 12, and 28), and as it comes back down (questions 10, 13, and 29).  According to Thornton and Sokoloff, student responses are only considered correct when all three questions within a given sub-cluster are answered correctly.\cite{thornsok} We expand on this point below.

\subsubsection{A generalized force-in-direction-of-motion model}

As with the Force Sled cluster, the Reversing Direction cluster provides possible responses that correspond directly with the $F \propto v$ model (or $a \propto v$ model for questions 27--29). For the questions in the Reversing Direction cluster, however, it is beneficial to consider a generalization of the $F \propto v$ model: the force/acceleration-in-the-direction-of-motion model.  This more general model ignores the magnitude of the force or acceleration throughout each part of the motion and only describes the direction.  Consider questions 11--13 (coin toss force questions).  A student using the $F \propto v$ model would indicate that the force on the coin is upward and decreasing as the coin goes up, zero at the top of its motion, and downward and increasing as the coin comes back down.  But what if a second student thinks that the force is upward and constant while the coin travels up, but agrees with the first student on the other two questions?  This student cannot be considered as using the $F \propto v$ model, but may be classified within the direction-of-motion model.  In fact, our $F \propto v$ student may also be categorized as using the direction-of-motion model.  In this way the direction-of-motion model allows a broader classification of students who have similar, but not necessarily identical ideas.  

Our decision to used the generalized direction-of-motion model is data-driven.  Research conducted by one author (T.I.S.) suggests that considering the direction-of-motion model for the Reversing Direction cluster allows many more students to be classified into a common model than the $F \propto v$ model.\cite{smiththesis2007}  Incorporating the direction-of-motion model into the results from the Force Sled or Force Graphs clusters, however, did not add any significant information.  We suspect this is due to the fact that only the Reversing Direction cluster includes scenarios in which an object moves in more than one direction during a single described motion.  Moreover, the Reversing Direction scenarios do not provide information as to how the speed of the object changes throughout its motion.  

Table \ref{fmce_vert} shows the responses that corresond with each of the models described above as well as the most common student responses. The most common student responses correspond directly with the responses indicating the use of one of our described models.  We note that Thornton only provided answers that indicated the direction of the force, not its magnitude.\cite{thorntonblue}  As such, there is no way to tell from his data the likelihood that a student used the $F \propto v$ model.  Also, Thornton's work only looked at the questions pertaining to the force on the object.\cite{thorntonblue}  His results, however, were replicated by one author (T.I.S.) for questions 27--29 asking about the object's acceleration.\cite{smiththesis2007}

\begin{table*}[hbt]
\caption[FMCE: Reversing Direction cluster]{The ``Reversing Direction" cluster on the FMCE.}
\begin{center}
\begin{tabular}{ccccc}
\hline\hline
\multicolumn{1}{p{1.75cm}}{\centering{Question
Number}}&\multicolumn{1}{p{3.25cm}}{\centering{Constant
Downward Force or Acceleration}}&\multicolumn{1}{p{3.5cm}}{\centering{Force or
Acceleration in the Direction of
Motion}}&\raisebox{-1.25ex}[0pt]{~~$F,a \propto v$~~}&\multicolumn{1}{p{3.5cm}}{\centering{Most
Common Student
Response\cite{thorntonblue}}}\\
\hline
8--9--10&a--a--a&(e, f, or g)--d--(a, b, or c)&g--d--b&(e, f, or g)--d--(a, b, or c)\\
11--12--13&a--a--a&(e, f, or g)--d--(a, b, or c)&g--d--b&(e, f, or g)--d--(a, b, or c)\\
27--28--29&a--a--a&(e, f, or g)--d--(a, b, or c)&g--d--b&g--d--b\footnote{The most common responses for questions 27--29 can be found in Ref. \cite{smiththesis2007}.}\\
\hline\hline
\end{tabular}
\end{center}
\label{fmce_vert}
\end{table*}%

\subsubsection{False positives in vertical toss situations}

We return to the point of requiring that students answer all three questions correctly (responses a--a--a within each question triplet). Consider the pattern of ``a--d--a" responses on a given question triplet.\cite{smiththesis2007}  This student might believe that a constant downward force is exerted on the object while it is moving both upward and downward, but that no force is exerted while the object is ``stopped" at the apex of its motion.  This line of reasoning may come from difficulties distiguishing between instantaneous velocity and change in velocity (as it relates to acceleration).\cite{accel}  The student may use the reasoning that since the ball has zero velocity, it is not moving; therefore, the acceleration is zero and the force exerted on the object must also be zero by Newton's second law.  For all of these reasons it is widely accepted that responses to the questions in the Reversing Direction cluster must be examined in conjunction with one another rather than individually, otherwise a student with serious problems understanding direction reversal will get 2/3 correct on this question triplet.

Furthermore, consider two students who give very similar incorrect responses: Student 1 answers ``g--d--b" (consistent with $F \propto v$) while Student 2 answers ``g--d--a." Student 2's ``a" response to the third question indicates a constant downward force or acceleration.\footnote{The $F \propto v$ model would require an increasing downward force for each of these questions.}  The more general direction-of-motion model accounts for both sets of responses, though. Again, it would be inappropriate to give Student 2 a 1/3 correct score, even though answer ``a" is correct. The correct response also distinguishes the $F \propto v$ model from the direction-of-motion model. We do not believe, though, that the direction-of-motion model is 1/3 more correct than the $F \propto v$ model! For this reason, we agree with Thornton and Sokoloff that a student should not be considered correct any part of a sub-cluster unless that student answers correctly on all three questions. We thereby avoid measuring false positives in the Reversing Directions cluster.

\subsection{Force Graphs Cluster}
The Force Graphs cluster (see figure \ref{force_graphs_ex}) asks questions about motions sometimes identical in physics content to those found in the Force Sled cluster, but differing in presentation. Students are provided with a description of the motion of a toy car and asked to select a graph that depicts the force exerted on the car.  All of the correct responses indicate that the applied force is either zero or nonzero and constant.  Table \ref{fmce_force} shows how responses to the questions in the Force Graphs cluster correspond with the various mental models as well as the most common student responses.

\begin{table}[ht]
\caption[FMCE: Force Graphs cluster]{The ``Force Graphs" cluster on the FMCE.}
\begin{center}
\begin{tabular}{cccc}
\hline\hline
\multicolumn{1}{p{1.5cm}}{\centering{Question Number}}&\raisebox{-1.25ex}[0pt]{$F \propto\frac{\Delta v}{\Delta t}$}&\raisebox{-1.25ex}[0pt]{$F \propto v$}&\multicolumn{1}{p{3.25cm}}{\centering{Most Common Student Response\cite{thorntonblue}}}\\
\hline
14&e&a&a\\
16&a&c&c\\
17&e&a, b&b\\
18&b&h&h\\
19&b&d&d\\
20&g&f&f\\
21&e&a, h&h, f, a\\
\hline\hline
\end{tabular}
\end{center}
\label{fmce_force}
\end{table}%

As with the Force Sled cluster, the most common student responses for the Force Graphs cluster correspond almost exactly with the responses indicating a student's use of the $F \propto v$ model.  We separate the clusters to observe if students master the content of one cluster before the other.  Research suggests\cite{tugk,kohl,kohl2} that students do not display as much knowledge when working with graphs as when using descriptive language, and data from the FMCE support the separation of questions into two clusters.\cite{thorntonwhite}  The answers to the physics depend on the context and format of the question. Philosophically, this supports the use of a resources framework, which can account for differences in reasoning based on contextual and representational differences in resource activation.

Note that in table \ref{fmce_force} we designate multiple responses for a single model on question 21.  Students are asked about the force on a car once it has been released (after being pushed).  On one hand, response ``a" seems to fit perfectly with the $F \propto v$ model (see figure \ref{force_graphs_ex}): the car moves at a constant velocity so a constant force must be applied.  On the other hand, what if the students don't ignore friction (though they are explicitly told to do) or use an ``impetus/force dies away" model?  In each of these cases, the car would slow down at a (perhaps) steady rate, indicating a positive yet decreasing force (response ``h").  Both of these responses can be considered a $F \propto v$ model in the sense of the need for a ``maintaining agency," with response ``h" including the use of the ``dying away" resource.

Also consider question 17, where research by one author (T.I.S.) has shown that some students who primarily use the $F \propto v$ model will choose response ``a"  instead of ``b" (the $F \propto v$ response).  Response ``a" is not entirely different from ``b" as it is congruent with the $F \propto v$ model in magnitude but not direction.  This may correspond to a confusion between left and right as negative and positive, indicating a difficulty with coordinate systems rather than with forces.  As such we have decided to categorize response ``a" to question 17 as indicative of the $F \propto v$ model if the student displays use of that model in other responses to this cluster.  Strong empirical evidence and the fact that question 17 is one of only two questions in the Force Graphs cluster to describe a motion with constant velocity heavily influenced our decision to consider response ``a" as corresponding to the $F \propto v$ model.  

\subsection{Acceleration Graphs Cluster}
The Acceleration Graphs cluster is similar to the Force Graphs cluster.  Students are asked about the acceleration of a toy car undergoing various types of motion.  Again, students must choose the graph they believe best represents the acceleration of the car for each scenario.  It should be noted that the parenthetical reminders of ``(constant acceleration)" that are found in the questions of the Force Graphs and Force Sled clusters are omitted from these questions.  We examine the Acceleration Graphs cluster from the perspective of students' difficulties distinguishing between the concepts of acceleration and velocity reported by Trowbridge and McDermott\cite{accel} to form a type of acceleration-proportional-to-velocity ($a \propto v$) model. The most common student responses shown in table \ref{fmce_accel} correspond closely with the $a \propto v$ model.  

\begin{table}[ht]
\caption[FMCE: Acceleration Graphs cluster]{The ``Acceleration Graphs" cluster on the FMCE.}
\begin{center}
\begin{tabular}{cccc}
\hline\hline
\multicolumn{1}{p{1.5cm}}{\centering{Question Number}}&\raisebox{-1.25ex}[0pt]{$a \propto\frac{\Delta v}{\Delta t}$}&\raisebox{-1.25ex}[0pt]{$a \propto v$}&\multicolumn{1}{p{3.25cm}}{\centering{Most Common Student Response\cite{thorntonblue}}}\\
\hline
22&a&e&e\\
23&b&g&f\\
24&c&b&b\\
25&b&f&f\\
26&c&a&a\\
\hline\hline
\end{tabular}
\end{center}
\label{fmce_accel}
\end{table}%

\begin{figure}[ht]
\centering
\includegraphics[totalheight=3cm]{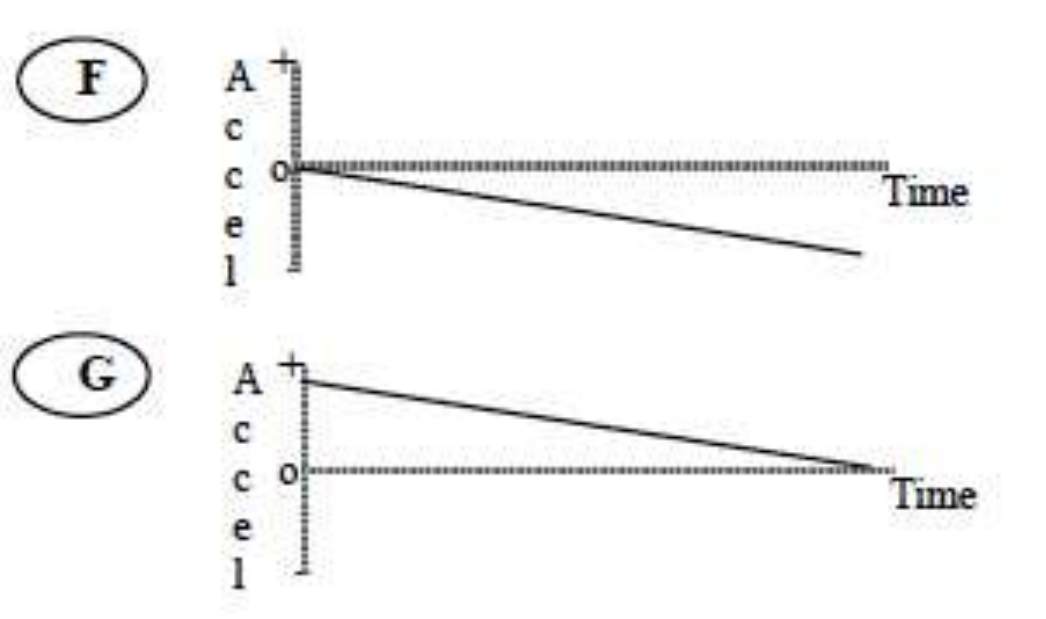}
\caption{Responses ``f" and ``g" for the Acceleration Graphs cluster.}
\label{accel_fg}
\end{figure}

We note a discrepancy between the $a \propto v$ model and the most common response for question 23.  This question asks students to choose the appropriate graph of acceleration vs. time for a car that ``moves toward the right(positive direction), slowing down at a steady rate."\cite{thornsok}  Figure \ref{accel_fg} shows responses ``f" and ``g" that correspond to the most common student response and the $a \propto v$ response, respectively.  Visually, responses ``f" and ``g" are incredibly similar, with identical magnitude slopes. They are also presented above one another (see figure \ref{accel_fg}).  Even though response ``f" would only accurately fit the $a \propto v$ model for a car moving to the left and speeding up (as in question 25), it is not surprising that students would choose ``f" for question 23.  Again, as with question 17, the problem may indicate issues with coordinate systems more than the relationship between acceleration and velocity.

\subsection{Newton III Cluster}
The Newton's third law cluster is the only cluster that commonly elicits two different incorrect student models: the mass dependence model and the action dependence model described above.  Table \ref{fmce_n3} shows how the responses to the questions in the Newton III cluster separate into these models.

\begin{table}[ht]
\caption[FMCE: Newton III cluster]{The ``Newton III" cluster on the FMCE.  *--Categorization for this response depends on the student's choice for question 36.}
\begin{center}
\begin{tabular}{ccccc}
\hline\hline
&&&&Most Common\\
\raisebox{1.5ex}[0pt]{Question}&\raisebox{1.5ex}[0pt]{Forces}&\raisebox{1.5ex}[0pt]{Mass}&\raisebox{1.5ex}[0pt]{Action}&Student\\
\raisebox{1.5ex}[0pt]{Number}&\raisebox{1.5ex}[0pt]{Equal}&\raisebox{1.5ex}[0pt]{Dependence}&\raisebox{1.5ex}[0pt]{Dependence}&Response\cite{smiththesis2007}\\
\hline
30&e&a&XX&a\\
31&e&a&b&e, f\\
32&e&a&b&b\\
34&e&XX&b&b\\
36&a&b&c&c\\
38&a&b*&b*, c&b\\
\hline\hline
\end{tabular}
\end{center}
\label{fmce_n3}
\end{table}%

The most common student responses shown in table \ref{fmce_n3} incorporate aspects of both the mass dependence and the action dependence models.\footnote{It should be noted that Thornton's study (Ref. \cite{thorntonblue}) did not include the questions from the Newton III cluster.  The most common student responses are those reported by one author (T.I.S.) in Ref. \cite{smiththesis2007}.}  We see that the most common set of responses shows more agreement with the action dependence model(questions 32, 34, 36, and 38) than the mass dependence model(question 30).  Response ``b" as the most common response for question 38 might seem a bit ambiguous (as it can be classified as either mass or action dependence), but one can use response ``c" for question 36 to then categorize both as uses of the action dependence model.  As with much of our analysis, this requires the assumption that students are reasoning relatively consistently from question to question.

\subsubsection{False positives in collision situations}

\begin{figure}[ht]
\centering
\includegraphics[totalheight=11.5cm]{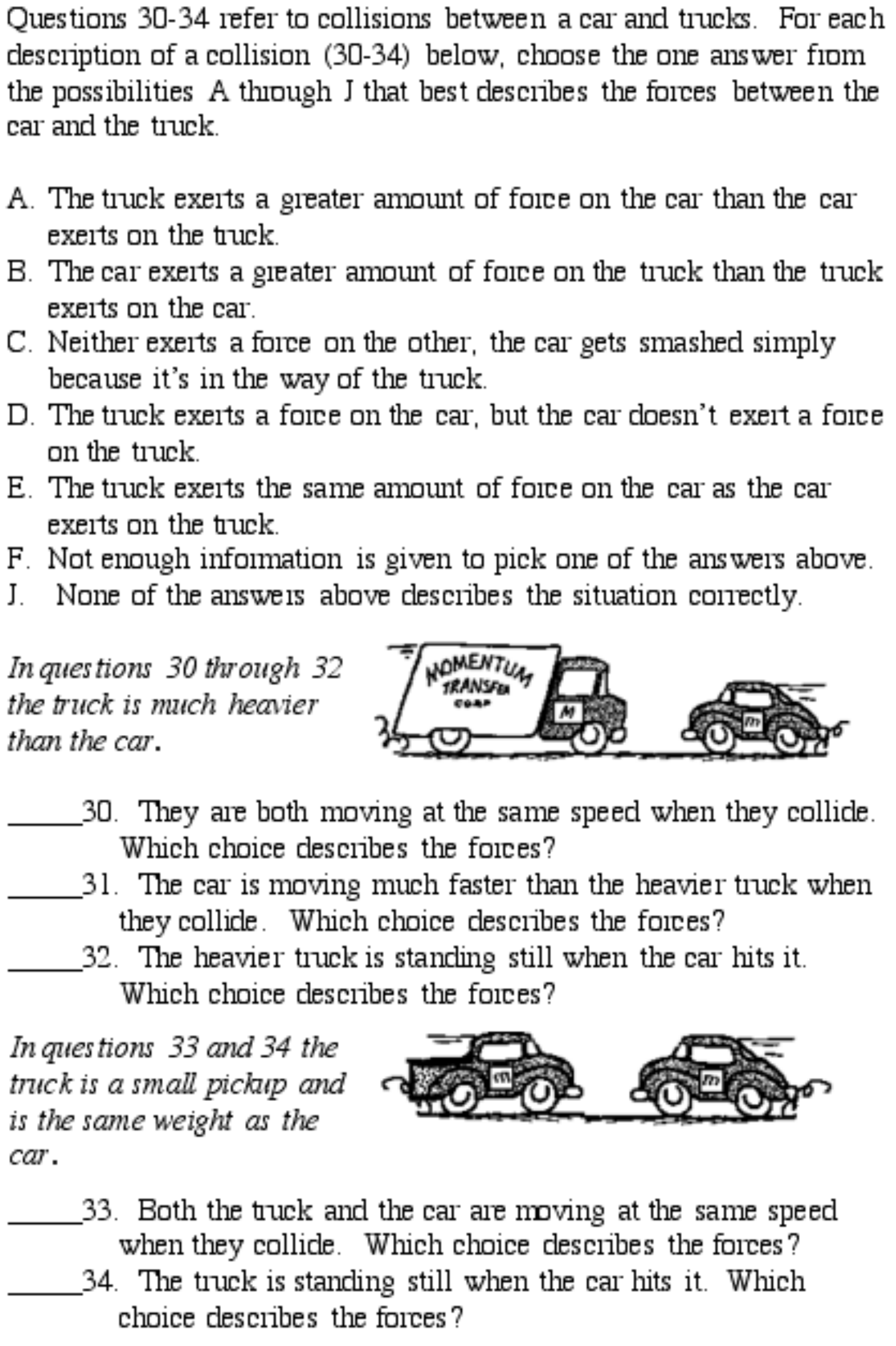}
\caption{Questions involving collisions within the Newton III cluster of the FMCE.  Please note that question 33 is not included within analysis of the FMCE.\cite{thorntonwhite}}
\label{n3_coll}
\end{figure}

The assumption of consistent reasoning has serious consequences when one considers question 31. Many students who answer ``e" on question 31 (the correct answer) answer incorrectly on questions 30, 32, and 34. \footnote{question 33 is not included in analysis of the FMCE.\cite{thorntonwhite}} (see figure \ref{n3_coll}).

\begin{figure}[ht]
\centering
\includegraphics[totalheight=4.1cm]{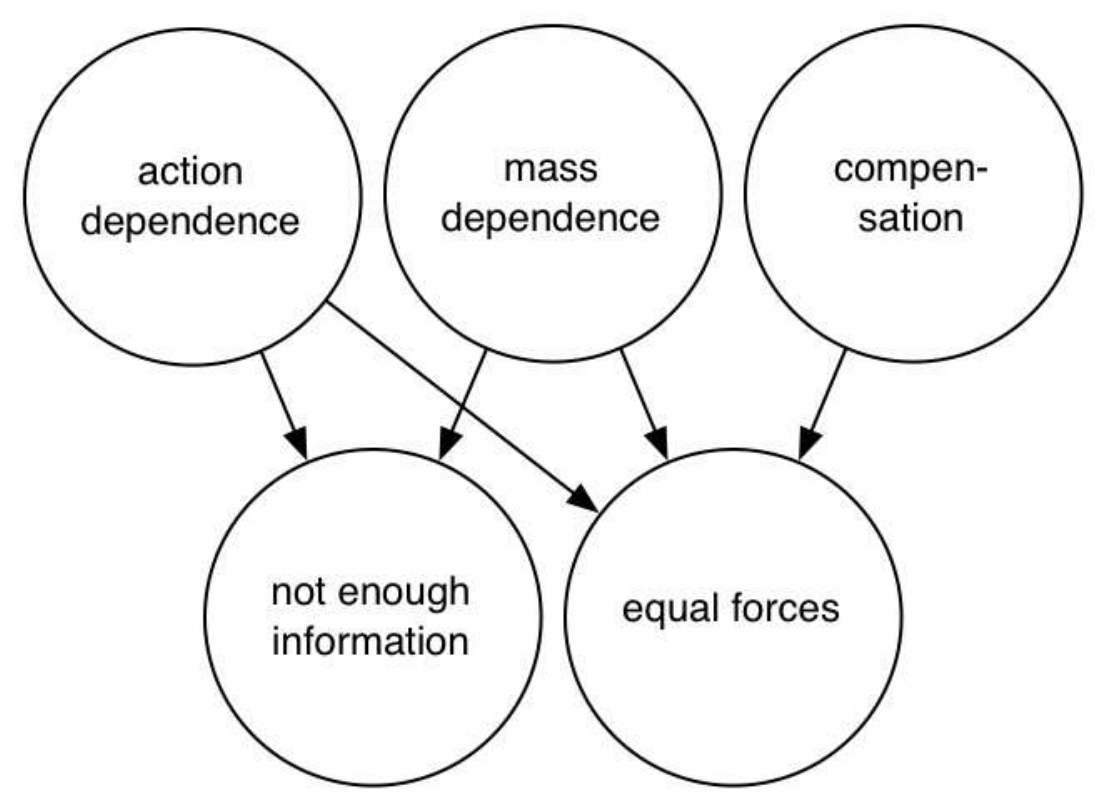}
\caption{The mass dependence and action dependence resources may be used together to cancel out some of their effects based on the situation (as in question 31).}
\label{cancel}
\end{figure}

We can infer with a fair degree of certainty why most students respond the way they do to questions 30 and 32.  In question 30, both vehicles are moving at the same speed before the collision (making action dependence a moot point), but the truck is much heavier than the car causing students to lean heavily toward mass dependence reasoning.  In question 32, the truck is still much heavier than the car, but it isn't initially moving.  As such, the greater ``activeness" of the car wins out and students use action dependence.  But what happens between these situations?  If the same two objects can interact in two different ways to get opposite results, there must be a situation in which the effects of mass dependence and action dependence will compromise or cancel out.  In question 31, the smaller, lighter car is initially moving much faster than the bigger, heavier truck, but the truck is moving.  In this case our ``most common student" must decide how to deal with mass dependence ideas from question 30 and action dependence ideas from question 32.  Response ``e" is one logical conclusion.  The two effects cancel each other out to result in the car and the truck exerting forces on each other that are equal in magnitude but opposite in direction.  A more discerning student, however, may feel that the effects will cancel each other to some degree but not necessarily completely, leading to response ``f," that more information is needed.  Figure \ref{cancel} shows how the mass dependence and action dependence resources may ``balance" to produce the correct conclusion.

To avoid the measurement of false positives, we suggest clustering responses 30--32 into a ``triplet" sub-cluster as done on the question triplet sub-clusters in the Reversing Directions cluster. Otherwise, one incorrectly rewards students for a situation where  two different wrongs do, in fact, make a right.

\subsubsection{False positives in pushing situations}

\begin{figure}[ht]
\centering
\includegraphics[totalheight=9.5cm]{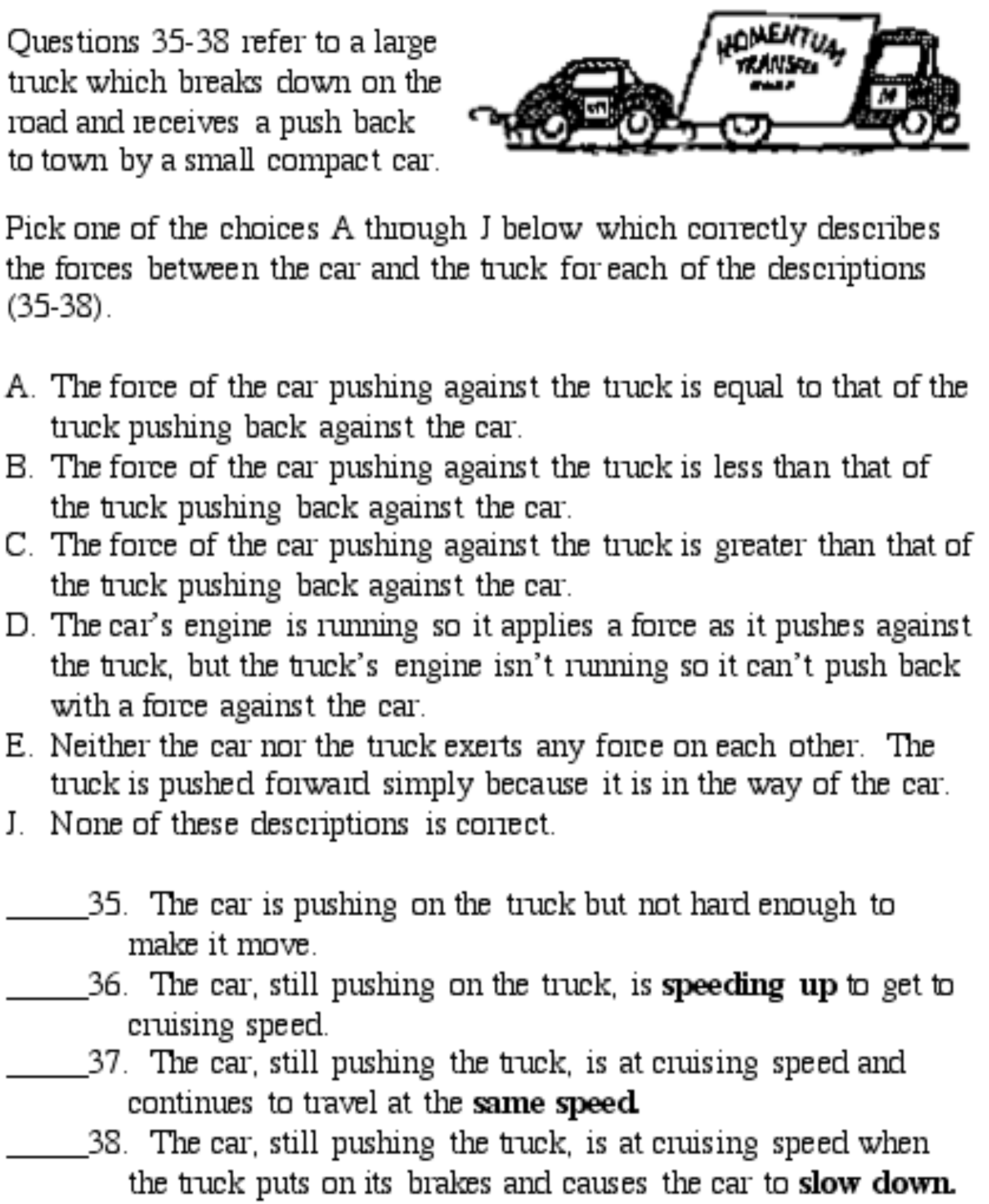}
\caption{Questions involving one vehicle pushing another within the Newton III cluster of the FMCE.  Please note that questions 35 and 37 are not included within analysis of the FMCE.\cite{thorntonwhite}}
\label{n3_push}
\end{figure}

A situation exists where two identical wrongs make a right, as well. As shown in table \ref{fmce_n3}, the most common responses for question 36 and 38 are ``c" and ``b," respectively (the questions are given in figure \ref{n3_push}).  Many students, however, choose responses ``c" and ``a" for these two questions.\footnote{These responses often occur with the most common responses listed in table \ref{fmce_n3} for questions 30--34.\cite{smiththesis2007}}  Response ``a" for question 38 indicates a correct answer of equal and opposite forces exerted by the two vehicles on each other.  Response ``c" for question 36, on the other hand, indicates the student's use of the action dependence line of reasoning. 

We again assume some consistency of student reasoning within a cluster of questions (that are contextually and representationally similar).  In question 36 the smaller car is pushing the heavier truck, and the two are speeding up.  Use of the action dependence resource suggests that the car is exerting a greater force than the truck (response ``c").  This result agrees with the pushing contextual feature reported by Bao, Hogg, and Zollman.\cite{baohoggzoll}  In question 38, however, the two vehicles are at a constant cruising speed, and the truck begins to apply its brakes, causing both vehicles to slow down.  Response ``b" for question 38 (the truck exerts more force) might be indicative of action dependence reasoning.  

When the truck begins applying its brakes, it may become the more active object in the student's mind, causing the vehicles to slow down. Once again, we have the possibility of two effects of incorrect resoning acting against one another.  The car is the active agent, pushing the truck forward. The truck is a second active agent, pushing back against the car. A student might believe that the two effects will perfectly balance each other and might arrive at the correct response (``a") in which the two vehicles exert equal amounts of force on one another.  This result is quite similar to that found for response ``e" on question 31 with the exception that question 38 presents the opportunity for students to use two conflicting versions of action dependence and eliminates the need for mass dependence.

\subsection{Velocity Graphs Cluster}
The Velocity Graphs cluster is very similar to the previously described ``graphing" clusters.  Students are presented with various descriptions of a car's motion, and they must choose the correct velocity vs. time graphical representation of the motion.  As with the Acceleration Graphs questions, the incorrect model we examine is derived from Trowbridge and McDermott's studies of students' understanding of kinematics and their difficulty distinguishing position from velocity.\cite{velocity}  This velocity/position confusion model is also closely related to Beichner's proposition that students view graphs as a picture of the situation no matter what the axes indicate.\cite{tugk}  Table \ref{fmce_vel} shows how the responses in this cluster correspond with the various student models.

\begin{table}[ht]
\caption[FMCE: Velocity Graphs cluster]{The ``Velocity Graphs" cluster on the FMCE.}
\begin{center}
\begin{tabular}{cccc}
\hline\hline
&Correct&Velocity/&Most Common\\
\raisebox{1.5ex}[0pt]{Question}&Model for&Position&Student\\
\raisebox{1.5ex}[0pt]{Number}&Velocity &Confusion&Response\cite{thorntonblue}\\
\hline
40&a&d&d\\
41&f&g&g\\
42&b&c, h&c\\
43&d&XX&XX\footnote{Thornton reports the most common student response for question 43 as ``not significant."\cite{thorntonblue}}\\
\hline\hline
\end{tabular}
\end{center}
\label{fmce_vel}
\end{table}%

\begin{figure}[ht]
\centering
\includegraphics[totalheight=3.5cm]{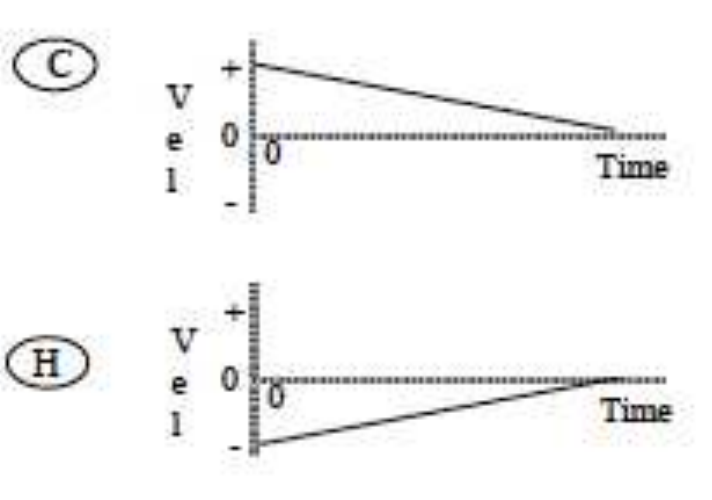}
\caption{Responses ``c" and ``h" for the Velocity Graphs cluster.}
\label{vel_ch}
\end{figure}

Once again, more than one response to a single question may indicate the use of our incorrect model.  On question 42, the toy car is said to be ``moving toward the left (toward the origin) at a steady (constant) velocity."  Responses ``c" and ``h" (shown in figure \ref{vel_ch}) both indicate a graph that gets steadily closer to the horizontal axis as time progresses.  For response ``c" a student could be picturing the car starting at the right and moving toward ``0," and students choosing ``h" could be triggered by the word ``left" to choose a graph that depicts negative velocity.  
 
As an aside, we note that questions 17, 23, and 42 form a cluster which lets one see if students have problems understanding coordinate systems, allowing for a finer grained analysis of students' understanding of force and motion. The FMCE only measures this topic implicitly, though, and the cluster is therefore relatively badly defined.

\subsection{Energy Cluster}
The Energy cluster on the FMCE contains questions that ask students to reason about the speed and kinetic energy of a sled after sliding down a hill.  The incorrect model for the questions in the Energy cluster, as described in section \ref{sec:resources}, corresponds with the idea that steeper hills will cause a greater change in speed and kinetic energy as the sled slides down.  Table \ref{fmce_energy} shows how the possible responses to questions in this cluster are divided among the correct and this incorrect model. The most commonly given incorrect answers correlate with the responses indicating a student's use of the slope dependent model.

\begin{table}[ht]
\caption[FMCE: Energy cluster]{The ``Energy" cluster on the FMCE.} \begin{center}
\begin{tabular}{cccc}
\hline\hline
&Energy/&Energy/&Most\\
Question&Speed&Speed&Common\\
Number&Depends on&Depends on&Student\\
&Height&Slope&Response\footnote{Most common student responses for the energy cluster discovered by research reported in Ref. \cite{smiththesis2007}.}\\
\hline
44&b&a&a\\
45&b&a&a\\
46&a&c&c\\
47&a&c&c\\
\hline\hline
\end{tabular}
\end{center}
\label{fmce_energy}
\end{table}%

\section{Summary}
\label{sec:summary} 
We have described a method for clustering questions on the FMCE that lets us use a resources framework to account for the most common student responses as reported by either Thornton\cite{thorntonblue} or in the work of one author\cite{smiththesis2007}. Our clustering allows us to categorize correct and incorrect responses using a single language of resource activation.  

Our clusters (Force Sled, Reversing Direction, Force Graphs, Acceleration Graphs, Newton III, Velocity Graphs, and Energy) take into account the physics content, the contextual aspects, and the representations used to ask the questions.  We show that students' incorrect responses to questions on the FMCE may be indicative of a variety of mental models that correspond with well documented research results. Using the resources framework, we can analyze sets of questions within some clusters (Reversing Direction and Newton III) to describe some correct student responses as false positives.  

We have presented interpretations for the most common incorrect responses for each questions, but this is in no way an exhaustive list of the possible mental models that may be used by students while answering questions on the FMCE.  Additional patterns of responses should be examined for prevalence among student responses and analyzed in terms of mental models that may be indicated by each.  For example, questions 17, 23, and 42 are a ``coordinate systems" cluster that has not yet been evaluated but may affect student responses on other clusters. Also, a second tier of mixed-context clusters of questions (such as Bao's\cite{baodiss}) could be created that ``slice" data in different ways. 

Research tools such as the FMCE are most effective to educators and researchers only when responses are examined to determine not only whether or not students have the correct ideas, but also what ideas they do have (correct or otherwise) and how consistently they use these ideas across similar questions. Our clustering allows such an analysis, giving insight into both how we model student thinking and how we could better address student needs in the classroom.

\begin{acknowledgments}
We thank three reviewers for their valuable comments and insights on an earlier version of this paper.  Supported in part by NSF grants DUE--0510614 and REC--0633951.  Portions of this manuscript originally published in Ref. \cite{smiththesis2007}.
\end{acknowledgments}
\bibliography{per}
\bibliographystyle{apsrev-stper}

\end{document}